\documentclass[twocolumn]{aastex62}
\hypersetup{linkcolor=magenta,citecolor=blue,filecolor=cyan,urlcolor=purple}

\usepackage{graphicx}
\usepackage[caption=false]{subfig}
\usepackage{ulem}

\received{2020 August 10}
\revised{2020 August 17}
\accepted{2020 August 17}
\submitjournal{ApJL}

\shorttitle{Magnetic Cloud Cross Helicity}
\shortauthors{Good et al.}

\begin{document}

\title{Cross Helicity of the November 2018 Magnetic Cloud Observed by the \textit{Parker Solar Probe}}

\correspondingauthor{S. W. Good}
\email{simon.good@helsinki.fi}

\author[0000-0002-4921-4208]{S. W. Good}
\affil{Department of Physics, University of Helsinki, Helsinki, Finland}

\author[0000-0002-4489-8073]{E. K. J. Kilpua}
\affil{Department of Physics, University of Helsinki, Helsinki, Finland}

\author[0000-0001-9574-339X]{M. Ala-Lahti}
\affil{Department of Physics, University of Helsinki, Helsinki, Finland}

\author[0000-0003-2555-5953]{A. Osmane}
\affil{Department of Physics, University of Helsinki, Helsinki, Finland}

\author[0000-0002-1989-3596]{S. D. Bale}
\affil{Department of Physics, University of California, Berkeley, CA, USA}
\affil{Space Sciences Laboratory, University of California, Berkeley, CA, USA}
\affil{The Blackett Laboratory, Imperial College London, London, UK}
\affil{School of Physics and Astronomy, Queen Mary University of London, London, UK}

\author[0000-0003-2555-5953]{L.-L. Zhao}
\affil{Center for Space Plasma and Aeronomic Research, University of Alabama, Huntsville, AL, USA}



\begin{abstract}

Magnetic clouds are large-scale transient structures in the solar wind with low plasma $\beta$, low-amplitude magnetic field fluctuations, and twisted field lines with both ends often connected to the Sun. Their inertial-range turbulent properties have not been examined in detail. In this Letter, we analyze the normalized cross helicity, $\sigma_c$, and residual energy, $\sigma_r$, of plasma fluctuations in the November 2018 magnetic cloud observed at 0.25~au by the \textit{Parker Solar Probe}. A low value of $|\sigma_c|$ was present in the cloud core, indicating that wave power parallel and anti-parallel to the mean field was approximately balanced, while the cloud's outer layers displayed larger amplitude Alfv\'enic fluctuations with high $|\sigma_c|$ values and $\sigma_r\sim0$. These properties are discussed in terms of the cloud's solar connectivity and local interaction with the solar wind. We suggest that low $|\sigma_c|$ is likely a common feature of magnetic clouds given their typically closed field structure. Anti-sunward fluctuations propagating immediately upstream of the cloud had strongly negative $\sigma_r$ values.

\end{abstract}



\section{Introduction}

The \textit{Parker Solar Probe} \citep[\textit{PSP};][]{Fox16} is now observing the solar wind closer to the Sun than any previous spacecraft. It seeks to establish how the corona is heated to $\sim$10$^6$~K temperatures, how the solar wind is formed and accelerated, how the wind observed in situ relates to coronal structure, and how the wind evolves with radial distance. Early findings include the discovery that short duration reversals in the radial component of the interplanetary magnetic field, previously observed by \textit{Helios} in fast wind at 0.3~au \citep{Horbury18}, are also a persistent feature of the near-Sun slow solar wind \citep{Bale19}; these `switchbacks' may, for example, be imprints of processes occurring in the solar atmosphere \citep[e.g., see the discussion in][]{Horbury20}, or they may arise in situ \citep{McManus20,Squire20}. It has also been found that the solar wind corotates with the Sun out to unexpectedly large radial distances \citep{Kasper19}.

During its first solar orbit in November 2018, at a heliocentric distance of 0.25~au, \textit{PSP} encountered a magnetic cloud originating from a coronal mass ejection (CME) on the far side of the Sun with respect to the Earth \citep{Korreck20}. A relatively slow-moving cloud, it displayed a complex magnetic flux rope structure \citep{Nieves20,Rouillard20} and likely accelerated solar energetic particles while closer to the Sun \citep{McComas19,Giacalone20}. Very few magnetic clouds have so far been observed by \textit{PSP} given that the solar activity cycle is currently at minimum and CME rates are low.

Magnetic clouds are the magnetically well-ordered, low plasma-$\beta$ subset of interplanetary coronal mass ejections (ICMEs) observed in situ \citep{Burlaga81}. Like ICMEs in general, they often travel faster than the ambient solar wind and expand as they propagate away from the Sun. Fast-mode waves generated upstream of fast-moving clouds may steepen to produce shocks. Downstream of shocks, the pile-ups of compressed and heated solar wind form sheath regions \citep[e.g.,][]{Kilpua17}. Sheaths and their magnetic cloud or ICME drivers are a major cause of geomagnetic activity \citep{Gosling91,Kilpua19}. Like the solar wind, the properties of magnetic clouds \citep[e.g.,][]{Bothmer98,Liu05,Wang05,Leitner07,Vrsnak19,Good19} and their sheaths \citep{Good20,Lugaz20} evolve with heliocentric distance.

The large-scale properties of magnetic clouds are reasonably well understood and have been extensively studied, in contrast to their small-scale properties. Like the solar wind, magnetic clouds display field fluctuations across a broad spectral range, with power spectra at frequencies below the ion gyrofrequency that are consistent with magnetohydrodynamic (MHD) turbulence theory \citep{Leamon98,Hamilton08}. The role of turbulence in heating magnetic cloud plasma has been investigated \citep{Liu06}. Localized regions of highly Alfv\'enic fluctuations within magnetic clouds have been identified \citep[e.g.,][]{Marsch09,Li16}, although their origins remain unclear.

In this study, we calculate the normalized residual energy, $\sigma_r$, and normalized cross helicity, $\sigma_c$, of MHD-scale fluctuations within the November 2018 magnetic cloud and surrounding solar wind. Values of $\sigma_r$ and $\sigma_c$ respectively indicate the degree to which fluctuations are Alfv\'enic, and the balance or imbalance of power in wave packets propagating parallel and anti-parallel to the mean magnetic field. The quantities are determined using a Morelet wavelet analysis similar to that applied by \citet{Chen13} in their study of the solar wind at 1~au. The wavelet technique gives a higher temporal resolution than can be accurately achieved with traditional Fourier analysis methods \citep{Torrence98}. In determining the temporal-spatial variation of $\sigma_r$ and $\sigma_c$, we seek to relate localized properties of the fluctuations to the global structure of the magnetic cloud and its interaction with the ambient solar wind, and to shed light on the question of whether the observed fluctuations were generated in the solar atmosphere or subsequently in interplanetary space. The near-Sun snapshot provided by \textit{PSP} allows fluctuations to be observed at a much earlier stage of development. It is possible that there was a greater solar imprint on fluctuations within this magnetic cloud at 0.25~au than in clouds observed further from the Sun. 

As part of their wider survey of small-scale flux ropes, \citet{Zhao20} analyzed $\sigma_r$ and $\sigma_c$ at low frequencies ($\sim$10$^{-5}$ -- $5\times10^{-4}$~Hz) during the magnetic cloud observation time. They found the cloud to be a magnetically dominated structure at these global scales, a finding that is consistent with the previous flux rope analyses of \citet{Telloni12} and \citet{Telloni13}. In contrast to the works of Zhao and Telloni, we analyze higher frequency fluctuations within the cloud that are at scales below the MHD turbulence outer scale and the system temporal scale size, i.e., the passage time of the magnetic cloud over the spacecraft.

\section{Spacecraft Observations}

\begin{figure*}
\epsscale{1}
\plotone{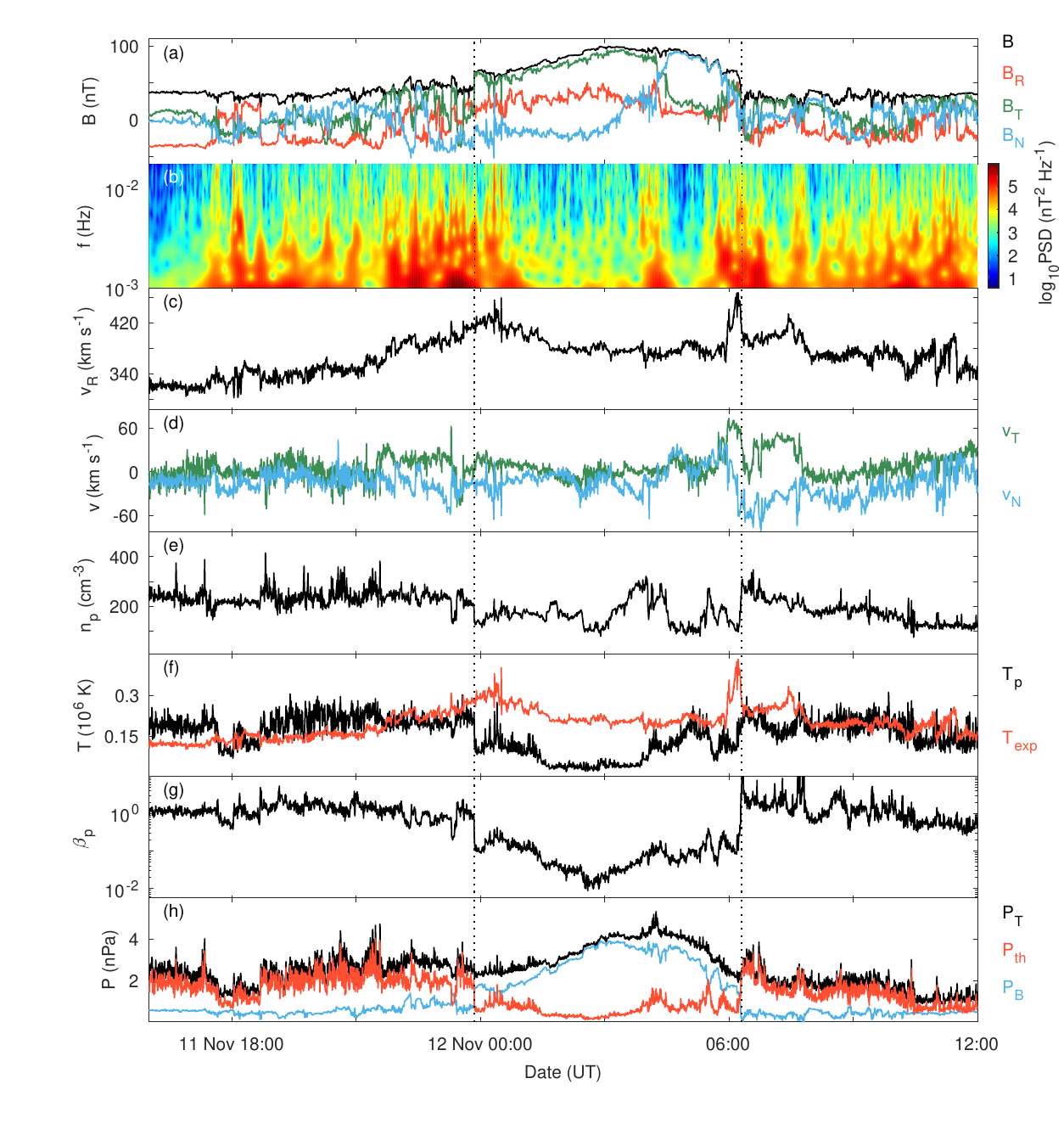}
\caption{\textit{PSP} observations of the magnetic cloud. The panels show the (a) \textbf{B} field in RTN coordinates, (b) wavelet PSD of the \textbf{B} field fluctuations sampled at inertial-range frequencies, (c) \textit{R} component of proton velocity \textbf{v}, (d) \textit{T} and \textit{N} components of \textbf{v}, (e) proton number density, (f) proton and expected temperatures, (g) proton plasma-$\beta$, and (h) the thermal, magnetic and total pressures. The cloud boundaries are shown with vertical dashed lines.}
\label{fig:Data}
\end{figure*}

Magnetic field data, \textbf{B}, from the FIELDS instrument suite \citep{Bale16} and plasma data from the SWEAP instrument suite \citep{Kasper16} on board \textit{PSP} have been analyzed. Figure~\ref{fig:Data} shows measurements from the instruments at a resolution of 27.96~s around the time of the magnetic cloud passage. The cloud boundaries, observed at November 11 23:51 and November 12 06:17~UT, are marked with vertical lines in the figure. The interval bounded by these lines displays all of the standard signatures of a magnetic cloud, including an enhanced \textbf{B} magnitude with relatively smooth large-scale variations in the \textbf{B} components, a proton temperature, $T_p$, lower than that predicted, $T_{exp}$, by the speed-temperature correlation relationship valid for non-cloud solar wind \citep{Lopez86}, and a proton plasma-$\beta\lesssim 0.1$. There is also a characteristic enhancement in total pressure, $P_T$, within the interval. The plasma thermal pressure plotted in Figure~\ref{fig:Data}, $P_{th}$, includes both the proton and electron contributions, the latter estimated by assuming an electron temperature of $T_e=2T_p$. This $T_e$ approximation is broadly consistent with the $\sim$20~eV electron temperature measured at \textit{PSP} around the time of the cloud passage using quasi-thermal noise spectroscopy \citep{Moncuquet20}. The magnetic pressure, $P_B$, is likely overestimated in the cloud interval because the magnetic curvature tension of the cloud's flux rope, which balances the pressure perpendicular to the \textbf{B} field, is not included in the pressure calculation \citep{Russell05}. A discontinuous feature in the field components was present in the rear half of the cloud, possibly representing a boundary between two smaller flux ropes \citep{Nieves20}. In the analysis that follows, we treat the entire cloud interval as a single, large-scale flux rope.

The mean proton speed within the cloud was $\sim$390~km~s$^{-1}$, somewhat higher than the $\sim$320~km~s$^{-1}$ speed of the unperturbed, upstream solar wind. However, the cloud was not propagating fast enough relative to the ambient solar wind to have driven a shock when observed by \textit{PSP}. Nor was the speed gradient immediately preceding the cloud, $\Delta v$, steepening to form a shock at the time of observation, since $\Delta v$ was less than twice the fast mode speed, $c_f$ \citep[e.g.,][]{Gosling86}. In this estimation, $\Delta v\approx 80$~km~s$^{-1}$ was taken as the difference between the speeds observed at approximately November 12 00:00 (just within the cloud) and November 11 21:00~UT (where there is a plateau in the upstream wind speed), and the mean value of $c_f\approx 110$~km~s$^{-1}$ was taken across the speed gradient; shock steepening would have required $\Delta v>2c_f$.

Figure~\ref{fig:Data}(b) shows a wavelet spectrogram of the trace power spectral density (PSD) of the \textbf{B} field in the frequency range 0.001--0.018~Hz. At these frequencies, which fall within the inertial range of MHD turbulence, there was generally higher fluctuation power in the solar wind than within the magnetic cloud. The highest power in the interval was observed immediately ahead of the cloud. Within the cloud, localized patches of enhanced power could be seen near the cloud boundaries and around the discontinuous feature to the cloud rear. The properties of these fluctuations are explored in further detail in the following section.

\section{Analysis}

\subsection{Magnetic Field Fluctuations}

\begin{figure}
\epsscale{1}
\plotone{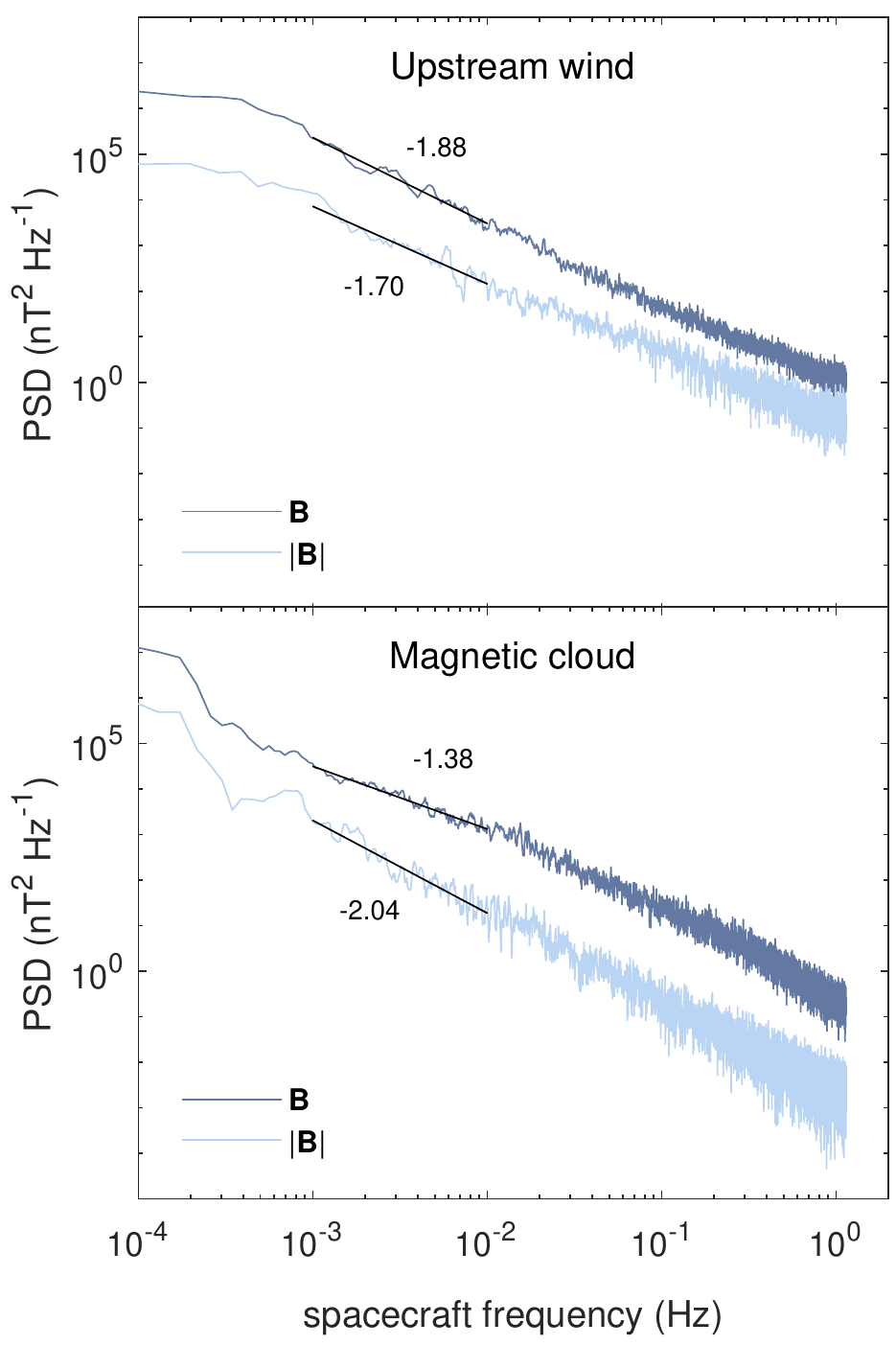}
\caption{Trace PSD of \textbf{B} fluctuations (dark blue) and PSD of \textbf{B} magnitude fluctuations (pale blue). Spectral indices in the $10^{-3}$--$10^{-2}$~Hz frequency range are indicated for each spectrum. The top panel shows spectra for the upstream wind from November 11 21:00~UT up to the cloud leading edge and the bottom panel shows spectra for the cloud interval.}
\label{fig:BSpectra}
\end{figure}

Figure~\ref{fig:BSpectra} shows the trace PSD of \textbf{B} field fluctuations (i.e., total \textbf{B} fluctuation power) and the PSD of compressive $|\textbf{B}|$ fluctuations in the interval from 21:00~UT to the cloud boundary and in the cloud interval. \textbf{B} field data at a resolution of 0.438~s were used to calculate the spectra. The frequencies shown are below the spacecraft-frame proton gyrofrequencies and ion inertial frequencies, with both $\gtrsim 3$~Hz in each interval. It can be seen that \textbf{B} and $|\textbf{B}|$ fluctuation power was generally greater in the upstream wind than in the cloud. Also, power in compressive $|\textbf{B}|$ fluctuations was a fairly small percentage of the total power in both intervals: for example, compressive power at $10^{-3}$--$10^{-2}$~Hz was $\sim$3.4$\%$ and $\sim$4.5$\%$ of total power in the upstream wind and cloud, respectively. Compressive power became relatively more significant at higher frequencies in the upstream wind but not in the cloud. The rest of this study focuses on the primarily incompressible fluctuations that were found in the $10^{-3}$--$10^{-2}$~Hz range. 

The inertial range spectral slopes in the cloud and disturbed upstream region differ somewhat from the $\sim f^{-1.6}$ power law that has been typically observed by \textit{PSP} in non-cloud solar wind around 0.25~au \citep{Chen20}. The -1.38 slope for the \textbf{B} fluctuations at $10^{-3}$--$10^{-2}$~Hz within the cloud is consistent with the results of \citet{Chen13}, who found that spectral slopes at 1~au are particularly shallow when fluctuation amplitudes normalized to the mean field, $\delta B/B$, are low, as is the case generally in magnetic clouds. Here $\delta B/B$ varied from 0.12 to 0.24 at $10^{-3}$--$10^{-2}$~Hz in the cloud (cf. $\delta B/B \sim 0.41-0.97$ in the upstream wind). At higher frequencies in the inertial range ($8\times10^{-3}$--$10^{-2}$~Hz), \citet{Hamilton08} also found shallower slopes in magnetic clouds compared to the ambient solar wind at 1~au. 

The mean correlation length of the three \textbf{B} components, $\lambda_B$, was estimated to be $3.3\times10^{6}$~km within the cloud, corresponding to a spacecraft frequency of $1.9\times10^{-5}$~Hz calculated with Taylor's hypothesis. These values, which are associated with the MHD outer scale, suggest that the $10^{-3}$--$10^{-2}$~Hz frequencies fell within the inertial range as previously assumed, and that the relative shallowness of the cloud spectral slope at these frequencies was unlikely to have been due to a broad transition between an $f^{-1}$ injection range and the inertial range. At $10^{-2}$--$10^{-1}$~Hz, the spectral slope of \textbf{B} fluctuations in the cloud steepened to -1.83; the upstream wind did not display this mid-range steepening, with a slope of -1.88 at $10^{-3}$--$10^{-2}$~Hz and -1.86 at $10^{-2}$--$10^{-1}$~Hz. The mean correlation length in the cloud was considerably longer than in the upstream wind, where $\lambda_B=4.1\times10^{5}$~km, equivalent to a spacecraft frequency of $1.4\times10^{-4}$~Hz.                     
\subsection{Cross Helicity \& Residual Energy}

\begin{figure*}
\epsscale{1.15}
\plotone{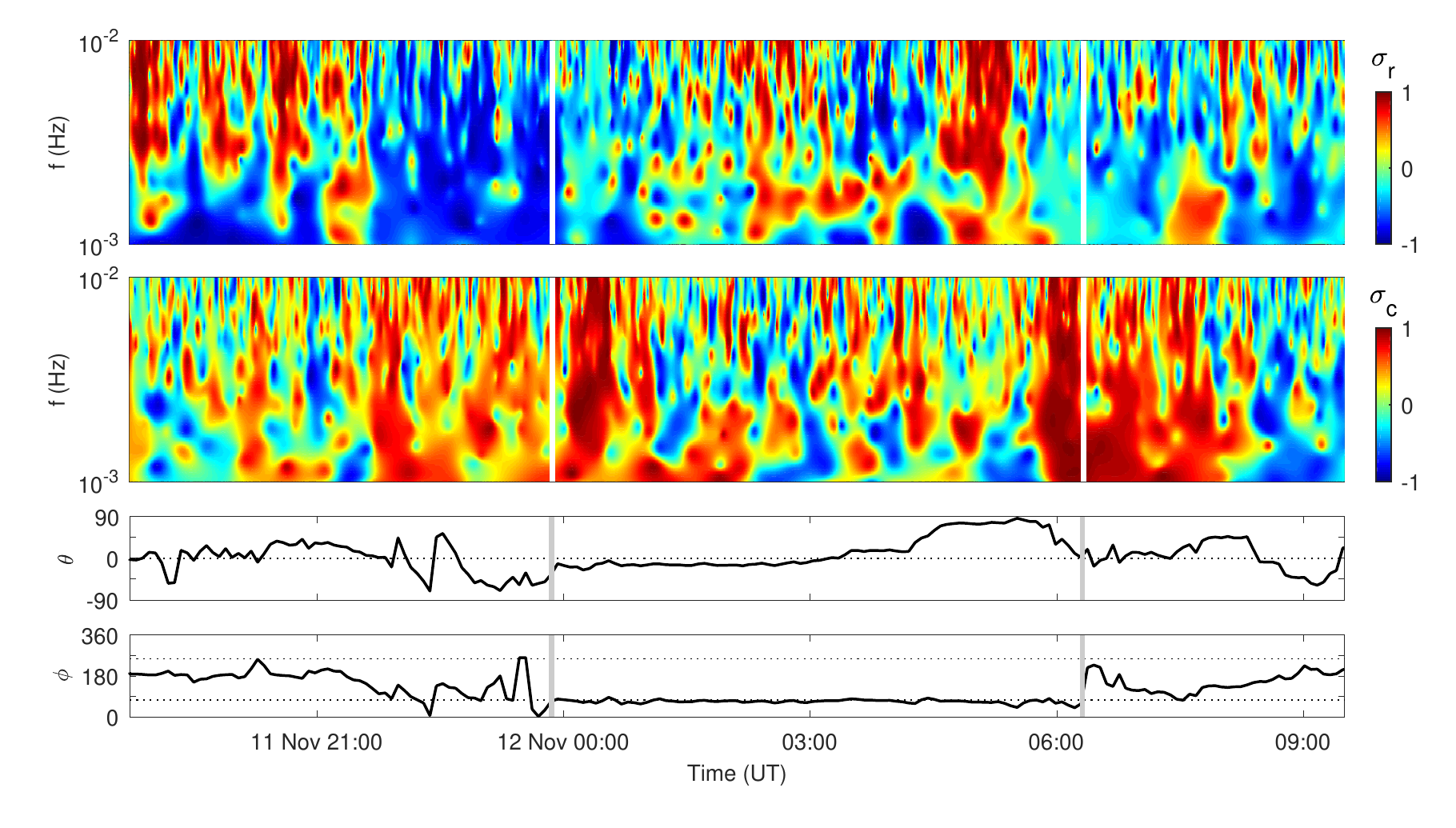}
\caption{Normalized residual energy, $\sigma_r$, normalized cross helicity, $\sigma_c$, \textbf{B} vector latitude angle, $\theta$, and \textbf{B} vector longitude angle, $\phi$. The \textbf{B} angles are with respect to the \textit{R-T} plane and indicate the mean field direction. Vertical lines demarcate the cloud interval. In the ambient solar wind, angles between the horizontal lines in the bottom panel correspond to anti-sunward vectors with respect to the local Parker spiral field.}
\label{fig:RECH}
\end{figure*}

Incompressible solar wind fluctuations may be treated as Alfv\'enic wave packets using the Elsasser variables, $\textbf{z}^{\pm}=\textbf{v}\pm\textbf{b}$, where \textbf{v} is the velocity, $\textbf{b}=\textbf{B}/\sqrt{\mu_0\rho}$ is the magnetic field in velocity units, and $\rho$ is the particle density. The $\textbf{z}^{+}$ mode corresponds to wave packets propagating anti-parallel to the background magnetic field and $\textbf{z}^{-}$ to packets propagating parallel to the field. The nonlinear interaction of $\textbf{z}^{+}$ and $\textbf{z}^{-}$ is the source of Alfv\'enic MHD turbulence in the solar wind.

The trace wavelet power spectra of \textbf{v}, \textbf{b}, and $\textbf{z}^{\pm}$, denoted by $E_v$, $E_b$, and $E_{\pm}$, respectively, may be used to define the normalized residual energy,
\[
\sigma_r=\frac{E_v-E_b}{E_v+E_b}
\]
and the the normalized cross helicity,
\[
\sigma_c=\frac{E_{+}-E_{-}}{E_{+}+E_{-}}.
\]
Values of $\sigma_r$ and $\sigma_c$ are limited to the range $[-1, 1]$, and $\sigma_r^2+\sigma_c^2\leq1$. Positive (negative) $\sigma_r$ values indicate an excess of energy in velocity (magnetic field) fluctuations, while values around zero indicate an equipartition of energy that is predicted for MHD Alfv\'en waves; positive (negative) $\sigma_c$ values correspond to wave packets propagating anti-parallel (parallel) to the background magnetic field being dominant, while values around zero indicate a balance of the parallel and anti-parallel fluxes.

Figure~\ref{fig:RECH} shows wavelet spectrograms of $\sigma_r$ and $\sigma_c$ across a similar time interval to that shown in Figure~\ref{fig:Data}. The spectrograms span the frequency range $10^{-3}$--$10^{-2}$~Hz, within the MHD inertial range. The mean proton number density value of 198~cm$^{-3}$ across the interval was used for the \textbf{B} normalization, with 4\% of the mass assumed to be from alpha particles. Vertical lines in the figure denote the magnetic cloud boundaries.

The bottom two panels indicate the mean \textbf{B} field direction. The latitude angle, $\theta$, gives the inclination of \textbf{B} relative to the \textit{R-T} plane, and longitude $\phi$ gives the angle between the projection of \textbf{B} onto the \textit{R-T} plane and the \textit{R} (anti-sunward) direction. The angles are calculated from successive 10 data-point ($\sim$4.7~min) averages of the field vector in order to approximate the background field direction. Values of $\phi$ between the horizontal lines (i.e., $75^{\circ}<\phi<255^{\circ}$) correspond to inward-directed field with respect to the Parker spiral field with a local, nominal spiral angle of 15$^{\circ}$, and $\phi$ values outside of this range correspond to outward-directed field. Reference to the Parker spiral is valid in the solar wind intervals but not within the cloud itself, where the background field is determined by the helical flux rope geometry.

\subsubsection{The Disturbed Upstream Wind}

From the start of the interval in Figure~\ref{fig:RECH} to November 11 21:40~UT, locally imbalanced patches of positive and negative $\sigma_r$ and $\sigma_c$ were present. This short subinterval was globally balanced, however, with mean values of $\langle\sigma_r\rangle=0.06$ and $\langle\sigma_c\rangle=0.05$. Between November 11 21:40~UT and the magnetic cloud leading edge time, $\langle\sigma_r\rangle$ fell to $-0.55$, indicating reduced Alfv\'enicity and the dominance of \textbf{B} fluctuation power, while $\langle\sigma_c\rangle$ rose to 0.39; since the mean field was primarily directed toward the Sun at this time, this positive cross helicity corresponded to wave packets propagating away from the Sun. The $\langle\sigma_r\rangle=-0.55$ value and corresponding -1.88 spectral slope (Figure~\ref{fig:BSpectra}) in this subinterval are in qualitative agreement with the correlation of negative residual energy with particularly steep spectral slopes identified at 1~au by \citet{Bowen18}. Some of the fluctuations in the vicinity of the cloud were likely generated by the magnetic cloud--solar wind interaction.

\subsubsection{The Magnetic Cloud \& Downstream Wind}

 Locally imbalanced $ \sigma_r$ and $\sigma_c$ were present throughout much of the magnetic cloud interval. Patches of strongly positive $\sigma_c$ immediately behind the cloud leading edge ($\sim$ November 12 00:30~UT) and preceding the cloud trailing edge ($\sim$ November 12 06:00~UT) are prominent features in Figure~\ref{fig:RECH}; these broadband Alfv\'enic fluctuations with $\sigma_r\sim0$ were located in the outer layers of the cloud's flux rope. The mean \textbf{B} field in both of these regions was oblique to the $\sim$\textit{R} propagation direction of the cloud and had a positive \textit{R} component, with $\phi\sim70^{\circ}$ in the front region and $\phi\sim60^{\circ}$ in the back region. Since $\sigma_c$ was strongly positive, the fluctuations propagated primarily anti-parallel to these mean field directions. The low $\sigma_r$ and high $|\sigma_c|$ regions in the cloud coincided with large-amplitude fluctuations in \textbf{B} and \textbf{v} seen in Figure~\ref{fig:Data}(a) -- (d). Global values across the cloud were $\langle\sigma_r\rangle=-0.03$ and $\langle\sigma_c\rangle=0.22$; excluding the high $|\sigma_c|$ outer layers (i.e., the first 40~min and last 30~min of the cloud interval) gives more globally balanced values of $\langle\sigma_r\rangle=0.02$ and $\langle\sigma_c\rangle=0.11$. 

The solar wind displayed moderately low $\langle\sigma_r\rangle$ (-0.16) and high positive $\langle\sigma_c\rangle$ (0.46) for a 2~hr period immediately following the cloud. The predominantly anti-sunward propagation of these Alfv\'enic fluctuations (i.e., toward the cloud) is consistent with them originating from the Sun rather than the cloud-wind interaction.

\section{Discussion}

\begin{figure}
\epsscale{1.15}
\plotone{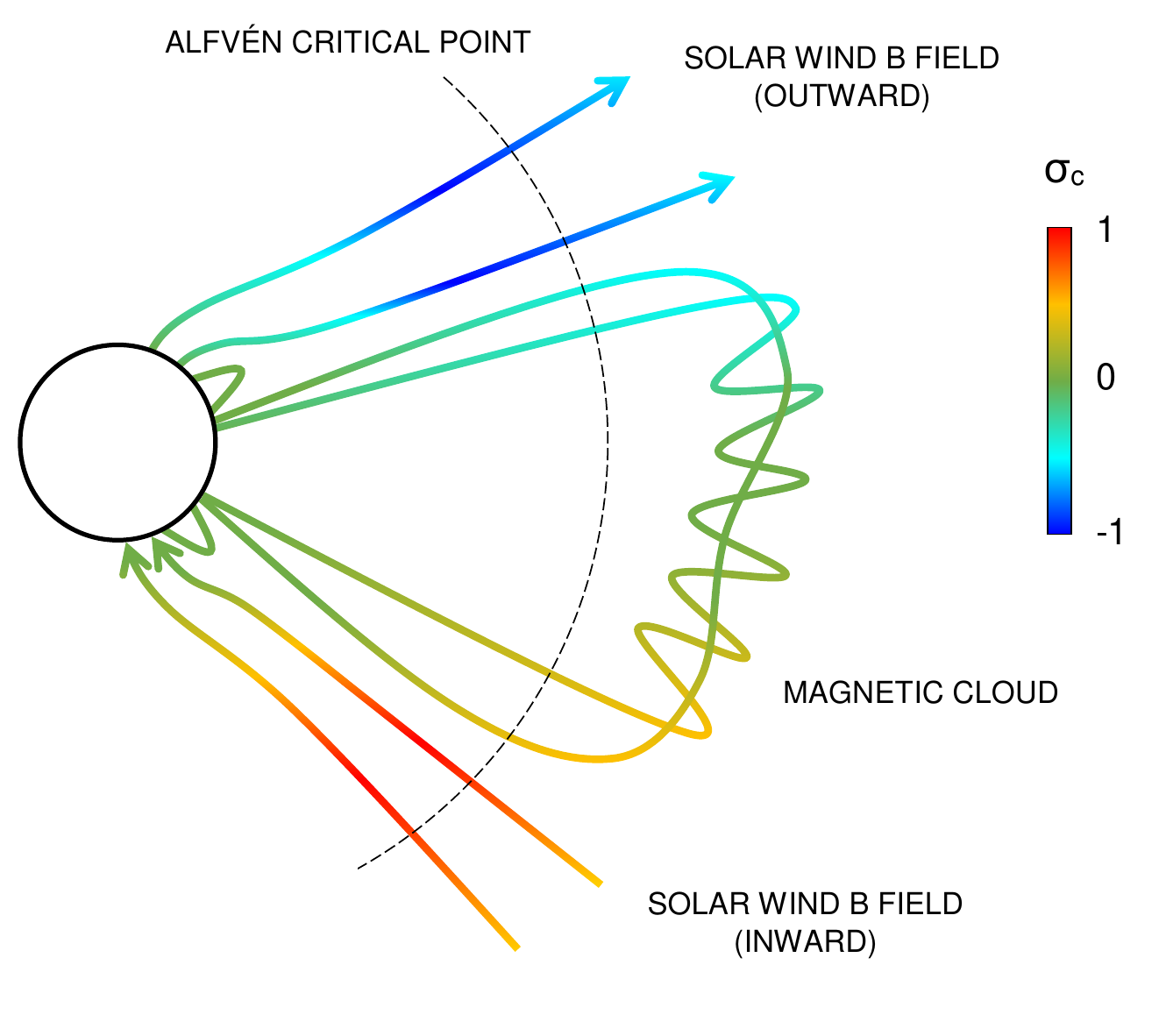}
\caption{A schematic picture of high $|\sigma_c|$ on solar wind field lines and low $|\sigma_c|$ in a closed-loop magnetic cloud beyond the Alfv\'en critical point. The radial distance of the critical point is not shown to scale. The reader is directed to the work of \citet{Zank18} for a quantitative model of $\sigma_c$ in the coronal magnetic field.}
\label{fig:CrossHScheme}
\end{figure}

The highly Alfv\'enic fluctuations in the magnetic cloud's outer layers are notable. At the cloud front, fluctuations may have been generated by the cloud's expansion and relatively fast propagation speed, with fluctuations propagating back into the cloud as well into the upstream solar wind. The high $|\sigma_c|$ in the cloud's outer layers suggests a dominant flux propagating away from a localized source of fluctuations, e.g., from the nose of the cloud pointing into the solar wind, or some other point along the cloud-solar wind interface where there was a large pressure gradient. If this interpretation is correct, penetration of the fluctuations into the cloud was limited to a relatively narrow outer layer. This may have been due to the mean \textbf{B} field being oblique to the normal of the interaction surface ($\sim$\textit{R} propagation direction of the cloud), with fluxes directed along the mean field being generated in preference to fluxes normal to the mean field. The cloud fluctuations could have been locally generated around the time of observation, or remnants of earlier interactions or processes that occurred closer to the Sun; the weakness of the cloud expansion (i.e., the low speed gradient across the cloud interval) and absence of strong interaction signatures at \textit{PSP} (e.g., a sheath with a shock) lend some weight to the latter possibility. The presence of Alfv\'enic fluctuations in the cloud's outer layers is consistent with the statistical analysis of clouds at 1~au performed by \citet{Li16}.

On the open field lines of the solar wind in the inner heliosphere, a dominant flux of anti-sunward Alfv\'enic fluctuations in the plasma frame is generally observed  \citep{Belcher71}. These fluctuations and a corresponding sunward component are thought to be generated in the corona below the Alfv\'en critical point. Any sunward fluctuations generated in this region propagate back to the Sun, leaving only the anti-sunward component to cross the critical point and be swept out with the solar wind. In contrast to solar wind field lines, the large-scale flux ropes found within magnetic clouds are generally thought to have both ends magnetically tied to the photosphere. `Sunward' and `anti-sunward' lose their distinction in this closed-loop case, and a mixed population of fluctuations with significant components that propagate both parallel and anti-parallel to the mean field may reach the critical point. This scenario, which is depicted in Figure~\ref{fig:CrossHScheme}, could explain the globally balanced cross helicity observed throughout much of the magnetic cloud. In contrast to the interplanetary origin outlined above, the high $|\sigma_c|$ fluctuations in the cloud's outer layers could have arisen if the outer field had reconnected with the surrounding magnetic field in the corona, giving it an open field topology that lead to the dominance of one flux component. The absence of bidirectional electron strahls in the trailing-edge layer \citep[][Figure 1]{Nieves20} supports this hypothesis, but their presence in the leading-edge layer indicates that the field here was connected at both ends to the Sun, like most of the cloud interval. A combination of balanced cross helicity in the cloud core arising in the corona and high $|\sigma_c|$ fluctuations in the cloud's outer layers arising from local interactions could also account for the in situ signatures at \textit{PSP}.

Low $|\sigma_c|$ is found in other situations. There is a tendency toward lower $|\sigma_c|$ in the solar wind with increasing heliocentric distance as in situ-generated sunward fluctuations develop, and lower $|\sigma_c|$ is often present in solar wind stream interaction regions \citep{Roberts87a,Roberts87b}. The low $|\sigma_c|$ observed before November 11 21:40~UT upstream of the November 2018 cloud may have been caused by the interaction of the solar wind with the cloud in interplanetary space. In contrast to the in situ origins for the above cases, we suggest that low $|\sigma_c|$ is a more intrinsic property of magnetic cloud plasma that is present within clouds at their earliest stages of existence close to the Sun. In agreement with our near-Sun case study, \citet{Hamilton08} reported lower inertial-range $|\sigma_c|$ values in magnetic clouds at 1~au compared to the fast or slow wind, and low $|\sigma_c|$ was also present in MHD modeling of a CME reported by \citet{Wiengarten15}.

We note finally that inferences from case studies are necessarily limited, and that a statistical study of $\sigma_c$ and $\sigma_r$ in magnetic clouds at 1~au is currently in preparation, which will allow broader conclusions to be made. Hopefully \textit{PSP} and the recently launched \textit{Solar Orbiter} will observe many more magnetic clouds \citep{Mostl20} in order to allow a comparable statistical picture to be produced for near-Sun heliocentric distances.

\section{Conclusion} 

We have analyzed the cross helicity and residual energy at inertial range frequencies ($10^{-3}$--$10^{-2}$~Hz) in a magnetic cloud at 0.25~au, the cloud closest to the Sun that has so far been observed in situ. Fluctuations immediately upstream of the cloud had negative residual energy and a positive cross helicity that corresponded to propagation away from the cloud. The magnetic cloud core had a fairly balanced global value of cross helicity ($\sigma_c=0.11$), indicating similar fluxes of Alfv\'enic wave packets propagating parallel and anti-parallel to the mean field direction. This may have been due to the cloud's flux rope being magnetically connected to the Sun at both ends, and the survival beyond the Alfv\'en critical point of a population of balanced $\sigma_c$ fluctuations originating in the corona. Given their closed field structure, magnetic clouds may have low $|\sigma_c|$ in general; we are unaware of any previous studies that have emphasized this point. The outer layers of the flux rope displayed highly Alfv\'enic fluctuations ($\sigma_r\sim 0$) with high $|\sigma_c|$, which may have been generated by local interaction between the cloud and solar wind or by the opening of the flux rope's outer field lines in the corona.


\acknowledgments

This work is dedicated to the memory of \mbox{Patricia} Ann Good (1949--2020). We thank the \textit{Parker Solar Probe} instrument teams for the data used in this study. Data were obtained from the CDAWeb archive (https://cdaweb.sci.gsfc.nasa.gov). S.G. and E.K. are supported by Academy of Finland grant 310445 (SMASH), and M.A.-L. and E.K. are supported by funding from the European Research Council (ERC) under the European Union’s Horizon 2020 research and innovation programme grant 724391 (SolMAG). S.G., E.K., M.A.-L. and A.O. are also supported by Academy of Finland Centre of Excellence grant 312390 (FORESAIL). We finally wish to thank the anonymous reviewer for their insightful and constructive comments on the manuscript.






\bibliography{bibliography.bib} 




\end{document}